\documentclass[useAMS,usenatbib]{mn2e}

\usepackage{graphicx}
\usepackage[T1]{fontenc} 
\usepackage{aecompl}

\RequirePackage{xspace}
\RequirePackage{amsmath}
\RequirePackage{amsfonts}
\RequirePackage{amssymb}

\def\ifundefined#1{\expandafter\ifx\csname#1\endcsname\relax}

\ifundefined{ensuremath}\def\ensuremath#1{\relax\ifmmode{#1}}
\else${#1}$\fi\else\relax\fi
\ifundefined{nuc}\def\nuc#1#2{\relax\ifmmode{}^{#1}{\protect\mathrm{#2}}
\else${}^{#1}$#2\fi}\else\relax\fi

\newcommand{\phx}{\texttt{PHOENIX}\xspace}
\newcommand{\nni}{\ensuremath{\nuc{56}{Ni}}\xspace}

\title[Feature of the $I$-band secondary maximum of a SN Ia]{Identification of the feature that causes the $I$-band secondary maximum of a type Ia supernova}
\author[D. Jack, E. Baron and P. H. Hauschildt]{D. Jack,$^{1}$\thanks{E-mail:
dennis@astro.ugto.mx} E. Baron$^{2,3}$ and P. H. Hauschildt$^{3}$\\
$^{1}$Departamento de Astronom\'\i{}a, Universidad de Guanajuato, Apartado Postal 144, 36000 Guanajuato, Mexico\\
$^{2}$Homer L. Dodge Department of Physics and Astronomy, University of Oklahoma, 440 W Brooks, Norman, OK 73019-2061, USA\\
$^{3}$Hamburger Sternwarte, Gojenbergsweg 112, 21029 Hamburg, Germany}
\begin{document}

\date{Accepted 2015 March 3.  Received 2015 February 16; in original form 2014 September 18}

\pagerange{\pageref{firstpage}--\pageref{lastpage}} \pubyear{2015}

\maketitle

\label{firstpage}

\begin{abstract}
  We obtained a time series of spectra covering the secondary maximum
  in the $I$-band of the bright Type Ia supernova 2014J in M82 with
  the TIGRE telescope.  Comparing the observations with theoretical
  models calculated with the time dependent extension of the \phx\
  code, we identify the feature that causes the
  secondary maximum in the $I$-band light curve.  
   Fe~II $3d^6(^3D)4s-3d^6(^5D)4p$ and similar high excitation
   transitions produce a 
   blended  feature at $\sim 7500\;$\AA, which  causes the rise of the
  light curve towards the secondary maximum.  
  The series of observed spectra of SN~2014J and archival data of SN~2011fe confirm
  this conclusion.
  We further studied the
  plateau phase of the $R$-band light curve of SN 2014J and searched
  for features which contribute to the flux. The theoretical models do not
  clearly indicate a new feature that may cause the $R$-band plateau
  phase.  However, Co~II features in the range of $6500-7000$~\AA\ and the
  Fe~II feature of the $I$-band are clearly seen in the theoretical spectra,
  but do not appear to provide all of the flux necessary for the
  $R$-band plateau. 

\end{abstract}


\section{Introduction}

The light curves of Type Ia supernovae  (SNe~Ia) are of general interest, in
particular for applications in cosmology. The optical
light curves of SNe~Ia admit a lightcurve shape relation
\citep{philm15,rpk96,philetal99,goldhetal01}  which allows them to be
used as ``correctable candles'' and hence, their distances can be
determined from the observed photometry.
The underlying physical explanation for the light curve shape
relation is that the total \nni mass produced in the explosion
determines to first 
order the temperature
and thus, more \nni leads to higher temperatures, which leads to
higher opacities, increasing the diffusion time 
\citep{nugseq95,kmh93,hkw95,PE00a,PE00b,PE01,mazzetal01}.  It has been
questioned whether the primary effect of the \nni mass is on the diffusion
time or rather on the color evolution of the spectra \citep{KW07}, but
the ultimate source for the observed luminosity of a SN~Ia is the
decay of \nni, and thus to zero order the luminosity should be
correlated with the mass of \nni.
Using the light curves of
SNe~Ia for distance measurements, led to the discovery of the phenomenon
that the Universe's expansion is accelerated, i.e, the discovery of
the dark energy
\citep{riess_scoop98,perletal99}. The nature of the dark energy is of
fundamental importance for cosmology.

Many physical details of Type Ia supernovae events are still not fully
understood \citep[see][for reviews]{hnar00,parrent14}. One of the most
important open questions is
the nature of the progenitor(s) of  SN Ia explosions.
The progenitor is believed to be a CO white dwarf in a binary system. Its companion 
star could be a main sequence star, an evolved star, or another white
dwarf. The primary white dwarf grows in mass via accretion, which can
be slow in the Chandrasekhar mass scenario, or rapid in scenarios that
involve the merging of two white dwarfs.
\citep[see][for reviews of the suggested possibilities]{levanon15,maoz14}.
In fact, SNe Ia may come from a variety of progenitor systems, and
detailed observations of nearby SNe Ia may lead to the understanding
of the nature of these progenitor systems \citep{nug_11fe_11,dilday11kx12}.
There is further
ongoing discussion 
about the explosion mechanism of Type Ia supernovae.
An instantaneous detonation in Chandrasekhar-mass WDs has
been ruled out, because it fails to produce intermediate mass elements (IME),
which are observed in the
spectra of SNe Ia \citep{arnett69}. In sub-Chandrasekhar-mass WDs
an instantaneous detonation can produce IME due to the lower
central densities 
\citep{shigeyama92,sim10}. However, the mass of the WD is rather
tightly constrained to produce both iron group elements and IME and the low densities
will not produce observed stable iron group elements.
Several different explosion models such as a pure deflagration,
delayed detonations, 
gravitationally confined detonations, violent mergers, and collisions have been suggested
\citep{nomoto84,khok91a,PCL04,jordangcd08,jordan12,pakmor12,pakmor13,rosswog09,kushnir13,long14},
which could 
explain the appearance of the intermediate mass elements observed in
SNe Ia spectra.
To understand the physical processes in a Type Ia
supernova further, different numerical codes have been developed and
used to model the spectra and light curves of SNe Ia
\citep{branch85,mazzali93,maeda14,wollaeger14,h03a,kasen06a,sim13,dessart10,jack11}.

We focus here on the light curves in the near-infrared wavelength
region.  The light curves of the near-infrared bands $I$, $J$, $H$, and $K$
very often show  a secondary maximum \citep{elias81,HGFS99by02}. In his
detailed study of near-infrared light curves of SNe Ia,
\citet{kasen06b} showed that the secondary maxima are caused by the
recombination of iron peak elements, which comes from a drastic
opacity change between ionization stages III to II of the iron
elements. This observation has been confirmed with other theoretical
radiative transfer calculations \citep{jack12,dessart14b,gall12}.
\citet{jack12} were able to assign  specific features of Fe II and Co
II that cause the individual secondary maxima in the $I$, $J$, $H$ and $K$
bands. However, their investigation was based only on theoretical light
curve calculations. \citet{dessart14b} found that [Co~II] lines play a
strong role, but this result has not been confirmed in other
calculations and may be due to the way that levels and superlevels are
coupled in their approximations.

In this work, we will present observational evidence that an Fe II
feature, produced by a blend of Fe~II lines, 
causes the secondary maximum in the $I$-band and additionally, we 
present an investigation of 
the plateau phase of the $R$-band light curve.
In \S~\ref{sec:methods}, we present the details of the observations
of SN~2014J and describe 
the methods used to calculate the theoretical light curves with the \phx\ code. In \S~\ref{sec:secmax}, we present the results
obtained for the secondary maximum in the $I$-band and the plateau phase
of the $R$-band.

\section[]{Methods}\label{sec:methods}

We obtained a time series of observations of high resolution spectra
of the bright SN~Ia,  2014J in M82 with the TIGRE
telescope.  We fitted the model light curves calculated with the \phx\
code to the observed light curves in the near-infrared $I$ and $R$-bands.
The observed and theoretical spectra are then
compared to determine the features that cause the secondary maximum in
the $I$-band and the plateau phase in the $R$-band.

SN~2014J was discovered by \citet{fossey14} in the nearby galaxy M82
approximately 1~week after explosion. We use for the explosion time
January 14.75 UT as found by \citet{zheng14}.

\subsection{Observations}

We observed SN 2014J in M82 with the TIGRE telescope situated in
Central Mexico close to the city of Guanajuato \citep{schmitt14}.  The
telescope TIGRE is a 1.2 m mirror telescope that operates completely
robotically. Its original design was to monitor the stellar activity
of solar-like stars, but other observational campaigns are also
feasible.  The TIGRE telescope is equipped with the HEROS spectrograph,
which has a high resolution $R \sim 20,000$ and covers the
wavelength range from about 3800 \AA\ to about 8800 \AA. The
spectrum is divided into two channels (red and blue).  The data
reduction pipeline is also fully automatic.

Beginning with the first observation during the night of 23rd of January 2014,
we were able to observe a spectrum of SN 2014J in M82 almost every
night following the discovery. See \citet{jack15} for a
complete presentation of the time series of SN~2014J spectra
observed with the TIGRE telescope.
We used an exposure time of 3 hours for every observed spectrum.
The spectra were obtained with the full spectral resolution $R \sim
20,000$. Since a supernova spectrum 
shows only very broad features and such a high resolution is not
necessarily required, we binned the spectrum down to a resolution of
10 \AA. 
The binning significantly improves the signal to noise ratio.
For this work, we are interested in the near-infrared wavelength range,
so that we will only use the spectra of the red channel (5800 \AA\ to
8800 \AA). SN 2014J showed 
significant reddening so that the obtained spectra in the blue channel
(3800 \AA\ to 5800 \AA) observed at such high resolution have a much
lower signal to noise.

\subsection{Theoretical light curves}

In order to model the theoretical light curves in different bands of
SN 2014J, we used the time dependent extension of the \phx\ code as
described in detail in \citet{jack09,jack11}. The code follows the
evolution of the supernovae envelope after the explosion enforcing energy
conservation. This includes energy deposition due to
$\gamma$-rays from the radioactive decay of $^{56}$Ni and $^{56}$Co.
Cooling due to the adiabatic expansion of the envelope, which is
assumed to be homologous, is also included. For the transport of
energy by radiation throughout the envelope, we solve the spherically
symmetric special relativistic radiative transfer equation
\citep{hauschildt92,hbjcam99}.
The envelope was divided into 128 layers. About 1,000 time
steps are performed for each point in the light curve.
We usually calculate a point in the light curve every 1 or 2 days.
All light curves have been calculated
assuming local thermodynamic equilibrium (LTE) and using about 2000
wavelength points.
The final spectra presented in this work have been calculated
using a higher resolution of about 20,000 wavelength points.
\phx uses the update Kurucz database \footnote{http://kurucz.harvard.edu/linelists/gfall/}
which contains about 80 million
atomic line transition.
 For the initial input structure, we used the
results of the W7 deflagration model \citep{nomoto84}, which also
includes the specific non-homogeneous abundances.

In this work, we focus on the secondary maxima in the
near-infrared wavelength region. We seek to obtain the best fit
to the individual observed light curves of SN 2014J in the respective
bands.  As described in \citet{jack12}, with the above assumptions, it
is necessary to
vary the equivalent two-level atom  thermalization parameter when solving the
radiative transfer problem. The source function of the radiative
transfer equation including scattering for an equivalent two level
atom can be written as
\begin{equation}
S_{\lambda}=(1-\epsilon_{\lambda})J_{\lambda}+\epsilon_{\lambda}B_{\lambda}.
\end{equation}
$S_{\lambda}$ is the source function, $B_{\lambda}$ is the Planck
function, and $J_{\lambda}$ is the mean intensity.  All these
quantities are wavelength dependent.  For the LTE \phx\ calculations,
it is possible to set a wavelength independent factor
$\epsilon=\epsilon_{\lambda}=$ constant to approximate LTE line
scattering over the whole wavelength range. We use this method here
for computational expediency.  Since the SN Ia envelope becomes
thinner during its free expansion phase, scattering becomes more and
more important, and the thermalization parameter $\epsilon$
decreases.  Applying this method of a decreasing thermalization
parameter, we obtained fits to the observed light curves of SN 2014J.

\section{Secondary Maximum}\label{sec:secmax}

We compare the spectra of SN~2014J observed with the TIGRE
telescope with the theoretical spectra calculated with \phx\ focusing on
the secondary maximum of the $I$-band light curve and the plateau phase
observed in the $R$-band.
All observed spectra and light curves have been dereddened with the values
$E(B-V)=1.33$ and $R_{\rm V}=1.3$ found by
\citet{amanullah14} in their study of the extinction law of SN 2014J.

\subsection{$I$-band}

We obtained a time series of spectra of SN 2014J during the
rise of the light curve towards the secondary maximum in the $I$-band.
Unfortunately, the HEROS spectrograph does not cover the whole range of
the $I$-band filter, meaning that we could not reconstruct a light curve
from our observed spectra. Therefore, we used the observed light curve
of the AAVSO database\footnote{www.aavso.org} to find the best fit of the model
light curve calculated with \phx. As described above, we varied the
thermalization parameter $\epsilon$ to obtain the best fit to the
observed light curve in the $I$-band.  Although this is not a physically
correct method to calculate a light curve, the overall trend of a
decreasing $\epsilon$ is physically motivated.  However, the goal here
is to use this
fit to study the spectral evolution of a SN Ia towards the secondary
maximum. 

\begin{figure}
\begin{center}
 \resizebox{\hsize}{!}{\includegraphics{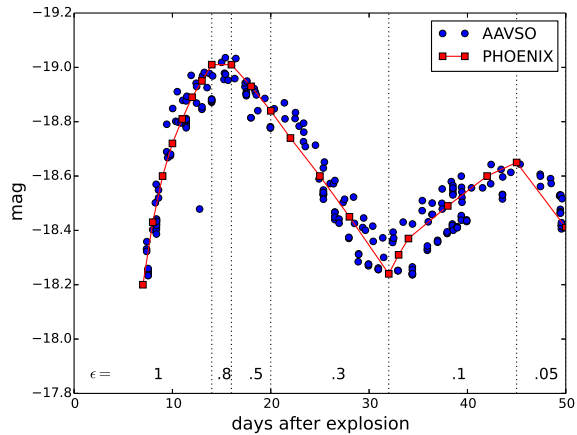}}
\caption{The circles represent the dereddened AAVSO observed light curve in absolute magnitude of SN 2014J in the $I$-band.
The solid line shows the best fit obtained from \phx\ models. 
The
values for $\epsilon$ are indicated along the 
bottom of the plot and the vertical dotted lines demarcate the range.}
\label{fig:lc_I}
\end{center}
\end{figure}

In Figure \ref{fig:lc_I}, the filled circles represent the light curve
of SN 2014J in the $I$-band obtained from AAVSO observation data.
The observed light curve has been dereddened.
We also plotted the AAVSO light curve in absolute magnitude using the distance
of 3.5 Mpc to M82, which results in a distance modulus of 27.7~mag,
which is comparable with the value of 27.6~mag found by \citet{foley14}.
The
solid line represents the best fit of the theoretical light curve
calculated with \phx\ models. 
As stated above, the models were calculated with different values
of $\epsilon$ for different epochs. This causes the kinks
in the model light curve.
The rise towards the secondary maximum
is clearly visible and starts around 32 days after the explosion.  The
secondary maximum is reached at around 45 days. These are the epochs
that are interesting for studying the spectral evolution and
identifying the feature that causes this secondary maximum.

\begin{figure}
\begin{center}
 \resizebox{\hsize}{!}{\includegraphics{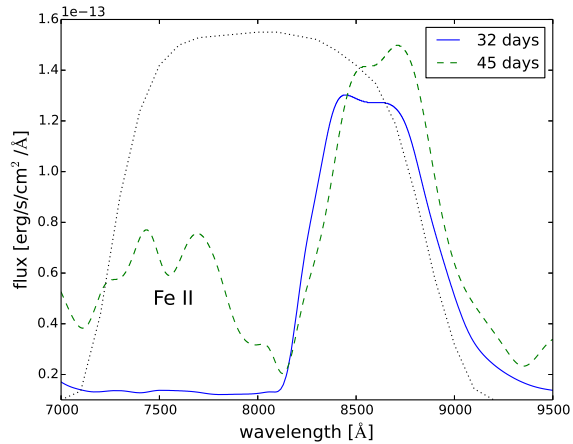}}
\caption{Theoretical spectra at different epochs after the explosion calculated with \phx.
At $\sim 7500$\;\AA, a new feature
of blended Fe II lines appears and causes the secondary maximum.
The dotted line indicates the $I$-band filter response.}
\label{fig:phx_spec_I}
\end{center}
\end{figure}

In Figure~\ref{fig:phx_spec_I}, we show a plot of the theoretical
spectra calculated with \phx\ at different epochs obtained from the best
fit to the observed light curve. 
The spectra are shown in flux received at Earth if the models would be at
the distance of SN 2011fe (6.6~Mpc) to be able to compare the fluxes directly.
The dotted line shows the $I$-band
filter response.   
At that epoch the spectrum does not show many clear
features.  The brightness decreases after the $I$ maximum. The 
Ca II IR triplet feature at $8500 - 9000$~\AA\ becomes
prominent. However, this feature,
 does not show a change in brightness during the rise to the
secondary maximum. From the beginning of the rise (32 days) up to the
secondary maximum (45 days) there appears a new feature of Fe II at  a
wavelength of $\sim 7500$ \AA.  
In Fe II there are numerous transitions from relatively high excited
states in this wavelength region, for example transitions from
$3d^6(^3D)4s-3d^6(^5D)4p$, $3d^5(^4P)4s4p(^3P^\circ)-3d^6(^5D)6s$, and
similar transitions, which have Einstein coefficient $A_{21}$ values $> 1\times 10^7$~s$^{-1}$.
Thus, this feature causes the observed
secondary maximum in the $I$-band of SNe Ia.

The observed spectra also clearly show the appearance of the  Fe II
feature mentioned above.
\begin{figure}
\begin{center}
 \resizebox{\hsize}{!}{\includegraphics{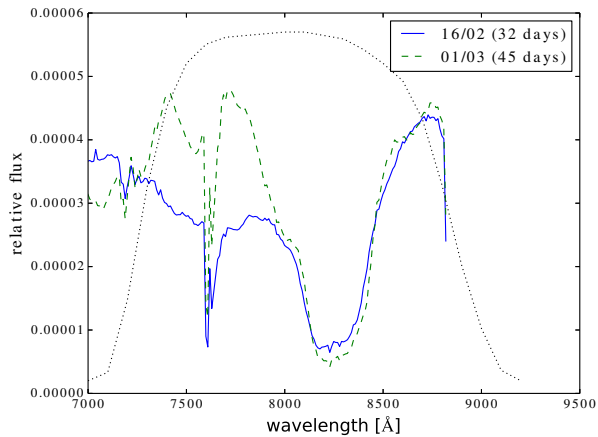}}
\caption{Spectra of SN 2014J observed on two different days before and during the secondary maximum. The appearance of the feature
of blended Fe~II lines can be seen in the observations at $\sim 7500$\;\AA\ as well.
The dotted line represents the $I$-band filter function.}
\label{fig:tigre_spec_I}
\end{center}
\end{figure}
Figure \ref{fig:tigre_spec_I} shows two observed spectra of SN 2014J
in the same wavelength range as was used for the theoretical spectra in Figure~\ref{fig:phx_spec_I}.
The dotted line represents the $I$-band filter response.
The spectrum from the 16th of February was taken directly before the
beginning of the rise towards the secondary maximum.
The spectrum from the 1st of March was observed during the secondary maximum. These observed spectra correspond to the theoretical
spectra of day 32 and day 45 in Fig. 2.
The feature of Fe II at around 7500 \AA\ also appears in the observed spectra. Note that there is a strong telluric absorption feature
around 7600~\AA, which contaminates the spectrum in this region.
However, we can clearly confirm that this Fe II feature actually
causes the secondary maximum in the $I$-band of SN~2014J, and thus, it
is likely the cause in most SNe~Ia.

Since our observed spectra of SN~2014J do not cover the whole wavelength
range of the $I$-band, we used archival data to obtain additional evidence
that the Fe~II feature causes the secondary maximum.

\begin{figure}
\begin{center}
 \resizebox{\hsize}{!}{\includegraphics{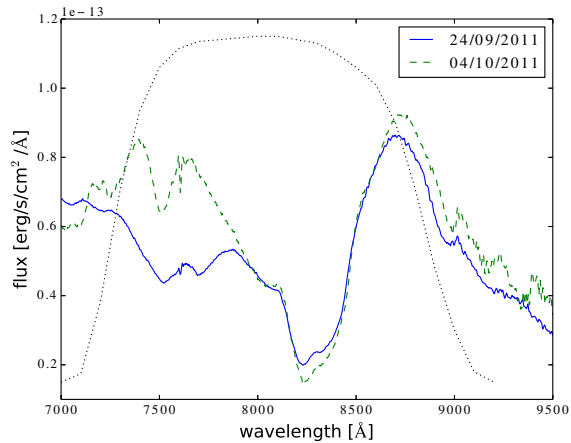}}
\caption{Spectra of SN 2011fe observed. The appearance of the feature
of blended Fe~II lines can be seen in the observations at $\sim 7500$\;\AA\ as well.
The dotted line represents the $I$-band filter function.}
\label{fig:sn2011fe_I}
\end{center}
\end{figure}

In Figure \ref{fig:sn2011fe_I} we show two spectra of SN 2011fe. The spectra were
observed by the Nearby Supernova Factory \citep{pereira13} in a time
series of spectra. The spectrum 
from the 24th of September 2011 coincides
with the minimum before the rise towards the secondary maximum in the $I$-band.
The secondary maximum was observed around the 8th of October
2011. Unfortunately, 
there does not exist a spectrum of that day. Therefore, we used the spectrum
of the 4th of October. In the plot one can clearly see that a feature
around 7500 \AA\ arises. This gives more observational
evidence that this Fe~II feature causes the secondary maximum
in the $I$-band of SNe~Ia.

\subsection{$R$-band}

The light curve of a typical type Ia supernovae in the $R$-band does not
show a secondary maximum, but usually a short plateau phase. Since all
other light curves in the optical wavelength range show a steady
decline after maximum light, there should also be a feature that
causes the effective $R$-band photosphere to ``hang up''.
We  used the same
method to identify this feature as in the study of the $I$-band
secondary maximum.  Using a varying line thermalization parameter
$\epsilon$, we obtained a best fit to the observed (dereddened)
$R$-band light curve of SN 2014J. 
\begin{figure}
\centering
 \resizebox{\hsize}{!}{\includegraphics{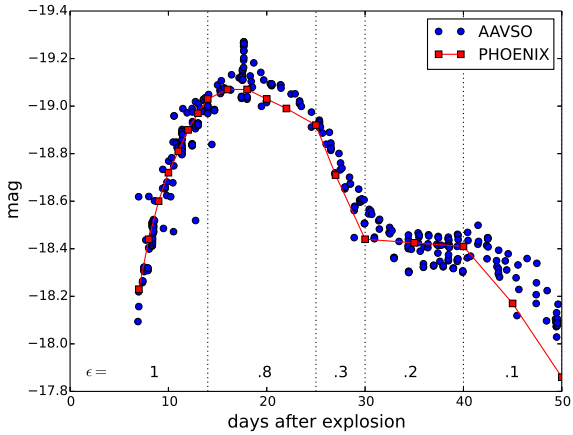}}
 \caption{
The circles represent the dereddened AAVSO observed light curve in absolute magnitude of SN 2014J in the $R$-band.
The solid line shows the best fit obtained from \phx\ models. The
values for $\epsilon$ are indicated along the 
bottom of the plot and the vertical dotted lines demarcate the range.}
\label{fig:lc_R}
\end{figure}

\begin{figure}
\begin{center}
 \resizebox{\hsize}{!}{\includegraphics{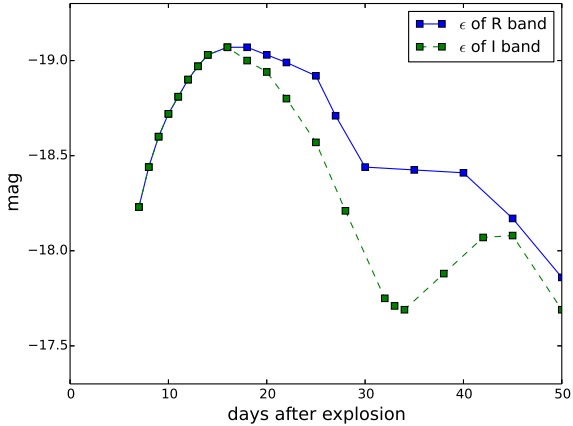}}
\caption{Here we show the \phx\ $R$-band light curve calculated with
the $\epsilon$ values for the $I$-band and $R$-band.}
\label{fig:compare}
\end{center}
\end{figure}

In Figure \ref{fig:lc_R}, the filled circles show the observed AAVSO
light curve in the $R$-band. 
Between about 30 and 40 days after  explosion the $R$-band
light curve of SN 2014J shows a short plateau phase, which we
also modeled with the theoretical light curve of \phx.
The solid line represents the best fit
obtained from \phx\ models by varying the thermalization
parameter. Note that the variations of $\epsilon$ are
slightly different from those used to compute  the $I$-band light
curve.
The differences, while small, represent a limitation in our
modeling approach. 
However, the overall trend of 
a decreasing $\epsilon$ is the same for all near-infrared bands.
Scattering becomes more and more important as the SN Ia envelope
expands.  

In Figure \ref{fig:compare} we compare the $R$-band
light curves that 
have been calculated using the values for $\epsilon$ 
used for the $R$-band and for the $I$-band.
One can see some clear differences during the phase of the minumum before
the plateau phase.
However, since $\epsilon$ is wavelength dependent it is not possible
to distinguish between a possible inaccuracy of the models or 
a real difference in the value of $\epsilon$ for the different
wavelength regions.

\begin{figure}
\begin{center}
 \resizebox{\hsize}{!}{\includegraphics{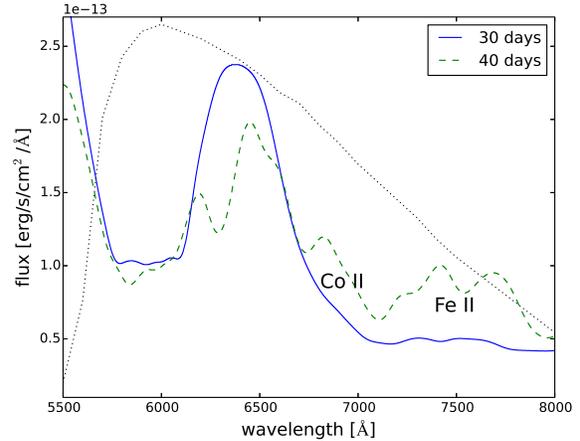}}
\caption{Spectra of the best fit to the $R$-band light curve.
Two features of blended Co~II lines and blended Fe~II lines contribute
to the flux that causes the $R$-band plateau phase.
The dotted line represents the $R$-band filter function.
The model luminosity has been scaled by the distance to SN 2011fe.}
\label{fig:phx_spec_R}
\end{center}
\end{figure}
We then studied the spectral evolution during this observed plateau
phase.  In Figure \ref{fig:phx_spec_R}, we show the theoretical
spectra calculated with \phx\ at the beginning of the plateau phase
(30 days) and at the end of the $R$-band plateau (40 days). 
The dotted line represents the $R$-band filter function.  After the $R$ maximum
(16 days), the brightness decreases in the whole wavelength range.  At
30 days, which marks the beginning of the plateau phase, the
characteristic P-Cygni Si~II feature around 6000 \AA\ to 6500 \AA\
strengthens. 
This feature changes between 30 and 40 days during the
plateau phase. However, the contribution of this feature to the brightness
in the $R$-band decreases and can, therefore, not cause the plateau phase.
In the models, blends of high excitation Co~II
lines appear in the wavelength region $6500-7000$~\AA, however the
total flux in the models is not large enough to produce the observed plateau.
The feature of Fe II at 7500~\AA\ that
causes the secondary maximum in the $I$-band is also present in the 
$R$-band filter, but again the filter sensitivity is falling off here so
the total flux is not large enough to produce the observed plateau. 

While nickel holes are seen in SNe~Ia \citep{fesen07}, the nickel hole
in W7 is too large to be compatible with galactic nucleosynthesis and
thus, the center of W7 has too much stable Ni and Fe, which will
reduce the total amount of Co. However, the Co~II lines seen in the
models are not easily identified in the observations (see below).

\begin{figure}
\begin{center}
 \resizebox{\hsize}{!}{\includegraphics{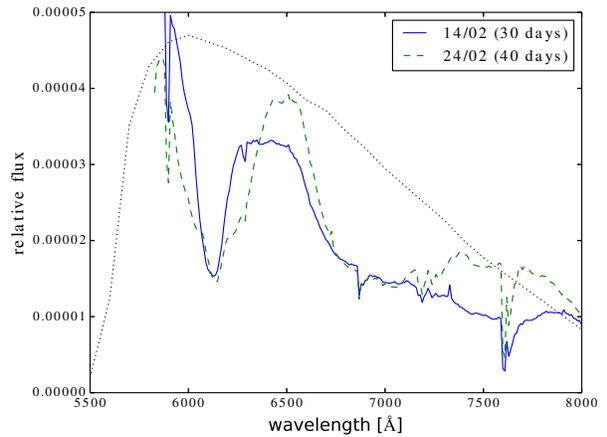}}
\caption{Spectra of SN 2014J observed on two different days at the beginning
and the end of the $R$-band plateau phase. The Fe~II features is clearly visible
while the Co~II feature of the models has not been identified.
The dotted line represents the $R$-band filter function.}
\label{fig:tigre_spec_R}
\end{center}
\end{figure}

Figure~\ref{fig:tigre_spec_R} shows the observed spectra of SN~2014J
in the $R$-band using the same wavelength range as for the theoretical
spectra of Fig.~\ref{fig:phx_spec_R}. The dotted line once again
represents the $R$-band 
filter function.  We plotted the spectrum from 14th of February, which
corresponds to the beginning of the plateau phase (30 days).  The end
of the plateau phase at 40 days is represented by the observed spectrum
from the 24th of February.  One can clearly see the change of the expansion
velocity of the characteristic Si~II feature as the photosphere moves
inwards.  There is no clear feature
arising in the $R$-band that obviously causes the short plateau phase.
Of course, the Fe~II feature at around 7500
\AA\ also contributes somewhat to the light curve in the $R$-band. As
discussed above Co~II lines in the $6500-7000$~\AA\ range are seen in
the models, but their signatures are not strong in the observed spectra.

\section{Conclusions}

We observed a time-series of spectra of the nearby Type Ia supernova
 2014J in M82 with the TIGRE telescope.  This time-series also
covered the secondary maximum in the $I$-band and the plateau phase in
the $R$-band.  We calculated the respective theoretical light curves
with the time dependent extension of the \phx\ code by varying the
line thermalization parameter to obtain a best fit to the observed light
curves.  For the $I$-band light curve of SN~2014J, we find that the
secondary maximum is caused by a blend of Fe~II high excitation lines because of
the recombination of iron peak elements from ionization stage III to
stage II. This confirms previous work  studying near-infrared
light curves of 
SNe Ia \citep{kasen06b,jack12,
  dessart14b}.  Studying the spectral evolution of SN~2014J, we find
observational evidence that the Fe~II feature at around
7,500 \AA\ causes the secondary maximum in the $I$-band. While W7 is
only a parameterized model of a real SN~Ia, the interior composition
is fairly generic due to the nuclear physics and thus, our conclusion
should be fairly general. However, the details are a bit more complicated,
and recent explosion calculations show
quite some diversity in the ejected material \citep{fink14,woosley11,moll14}.

As far as we know, there exists no detailed study of the $R$-band light
curves of Type Ia supernovae.  We observed SN~2014J, and it ---
like most other SNe Ia --- shows a short plateau phase in the $R$-band between
30  and 40 days after the explosion. We reproduced the observed
light curve with our \phx\ models to be able to identify the feature
that causes this short plateau phase. 
Co~II lines in the range $6500-7000$~\AA\ do increase in flux in the
models, and the Fe~II lines to the red may also contribute somewhat,
but the total flux from these features does not appear to be large
enough in the models and the Co~II signatures are not clearly seen in
the observations.

For future work, it will be very interesting to study the spectral
evolution of a Type Ia supernova in the other near-infrared bands $J$, $H$,
$Y$, and $K$. With this one might be able to identify the features which
cause the secondary maxima in these bands  as proposed
by \citet{jack12}. A further, more detailed study of the $R$-band is
important, since understanding the detailed composition and ionization
structure of the ejecta with time will help to constrain explosion models.

\section*{Acknowledgments}

We would like to thank the TIGRE team for having made possible the
observations of SN 2014J on short notice.  We acknowledge with thanks
the variable star observations from the AAVSO International Database
contributed by observers worldwide and used in this research. We thank
Kevin Krisciunas for helpful discussion on the $R$-band light curve.
The work has been supported in part by support for
programs HST-GO-122948.04-A provided by NASA
through a grant from the Space Telescope Science Institute, which is
operated by the Association of Universities for Research in Astronomy,
Incorporated, under NASA contract NAS5-26555.  This work was also
supported in part by the NSF grant AST-0707704.  This research
used resources of the National Energy Research Scientific Computing
Center (NERSC), which is supported by the Office of Science of the
U.S.  Department of Energy under Contract No.  DE-AC02-05CH11231; and
the H\"ochstleistungs Rechenzentrum Nord (HLRN).  We thank both these
institutions for a generous allocation of computer time.

\bibliographystyle{mn2e}
\bibliography{all}

\bsp

\label{lastpage}

\end{document}